\documentclass[12pt]{article}
\usepackage{amssymb,amsmath,amsthm,amsfonts,amscd}

 \usepackage{graphicx}
 \usepackage{float}
 \usepackage{caption}
\usepackage{subcaption}
\usepackage[font={small}]{caption}

\newcommand\be{\begin{equation}}
\newcommand\ee{\end{equation}}
\newcommand\bea{\begin{eqnarray}}
\newcommand\eea{\end{eqnarray}}

\newcommand{\fatalpha}{{\bf \alpha \kern -0.44em \alpha}}
\newcommand{\fatsigma}{{\bf \sigma \kern -0.54em \sigma}}
\newcommand{\tpchi}{{\bf D \kern -0.35em D}}
\newcommand{\llambda}{{\bf \lambda \kern -0.45em \lambda}}

\renewcommand{\theequation}{\arabic{equation}}
\renewcommand{\theequation}{\thesection.\arabic{equation}}
\bibliography{plain}

\title{\bf Quantum discord protection of  a two-qutrit V-type atomic system from decoherence by partially collapsing measurements
} \vspace{20mm}
\author{$\mathrm{N.\ Behzadi}^\text{1}$, $\mathrm{E.\ Faizi}^\text{2}$ and $\mathrm{O.\ Heibati}^\text{2}$
\\
$${\small $^1\mathrm{Research\ Institute\ for\ Fundamental\ Sciences, University \ of\ Tabriz, Tabriz, Iran.}$} \\$${\small $^2\mathrm{Department\ of \ Fundamental \ Sciences, Aazarbaijan\ Shahid\ Madani\ University, Tabriz, Iran.}$}}  \pagebreak

\vspace{20mm}

\begin{document}
\maketitle \vspace{15mm}
\newpage
\begin{abstract}
In this paper, by exploiting the weak measurement and quantum measurement reversal (WMQMR) procedure, we propose a scheme to show how one can protect the geometric quantum discord (GQD) of a two-qutrit V-type atomic system each of which interacts with a dissipative reservoir independently. We examine the scheme for the GQD of the initial two-qutrit Werner and Horodecki states for different classes of weak measurement strengthes. It is found out that the presented protocol enables us to suppress decoherence due to the amplitude damping (AD) channel and preserve the quantum discord of the two-qutrit system successfully.
\\
\\
 {\bf Keywords: Geometric quantum discord, Decoherence,
Weak measurement, Quantum measurement reversal. }
\end{abstract}

\newpage
\section{Introduction}
One of the most  prominent aspects of quantum mechanics is quantum correlation which plays a vital role in quantum information tasks \cite{computation}. For a long time, the only known source of quantum correlations in composite quantum systems has been  entanglement which is responsible for various quantum information processing tasks, such as, quantum cryptography \cite{crypto}, quantum teleportation \cite{teleport} and quantum dense coding \cite{dense}. Recently, it has been shown that entanglement is just one form of quantum correlations and some applications in quantum information theory do not require it. Oliver and Zurek in \cite{Zurek}, claimed that there is a new measure of quantum correlations other than entanglement. Despite entanglement, the new non-classical correlation which was named quantum discord can even exist in separable  states. Some tasks
that benefit from quantum discord are state merging \cite{state merging1, state merging2}, remote state preparation \cite{remote}, entanglement
distribution \cite{distribution1,distribution2}, transmission of correlations \cite{transmission} and information encoding \cite{information encoding}.

Quantum correlations may be corrupted by decoherence because the quantum system unavoidably coupled to its surrounding environment and lose coherence \cite{decoherence}.  Thus, protecting quantum correlations from  decoherence due to environmental noises is critical in quantum information theory. Several methods have been explored to suppress decoherence, including decoherence-free subspace \cite{decoherencefree1, decoherencefree2}, entanglement distillation \cite{distillation1, distillation2, distillation3}, quantum Zeno effect \cite{zeno} and quantum error correction \cite{errorcorrection1, errorcorrection2}. Additionally, WMQMR has been offered to deal with  decoherence due to the AD channel.

The concept of weak measurements was suggested by Aharonov $et$ $al$. \cite{aharanove} for the first time. Later on, by making use of weak measurement and its reversal, Katz $et$ $al$. presented a way to suppress the decoherence of amplitude damping for a single phase qubit \cite{katz}. Recently, it has been shown that fragile quantum states of a single-qubit system \cite{singlequbit1, singlequbit2, singlequbit3} and quantum entanglement of two-qubit systems \cite{twoqubit1, twoqubit2, twoqubit3} can be protected by WMQMR. Xiao $et$ $al$. applied this method to preserve the entanglement of a two-qutrit system from decoherence \cite{li}. Recently, it has been demonstrated in \cite{li} that WMQMR is suitable for protecting quantum discord from decoherence of the AD noise.
In a quantum measurement, by performing the so-called von Neumann measurement on a system, the state of concern instantaneously collapses to one of the eigenstates of the system. Since von Neumann measurement destroys the initial state, it is impossible to recover it from the outcome of the measurement. In contrast to this type of quantum measurement, there is a partial collapsing or weak measurement in which the state of the system does not totally collapse, such that the final state has some information of the original state. In this case, the exact initial state can be recovered by applying the reversed weak measurement operator.

Motivated by the above assumptions, our aim of this paper is the investigation of the role
of WMQMR in protecting the two-qutrit discord, measured by the geometric quantum discord, from destruction caused by the AD noise. We examine the performance of this method for the initial two-qutrit V-type Werner and Horodecki states. In \cite{threelevel}, the exact value of geometric discord for families of three-level states with maximally mixed marginals such as Werner state, Bell state and Horodecki state is evaluated, however, both of Werner and Horodecki states after undergoing the AD channel do not have maximally mixed marginals. Thus, instead of calculating the exact values of GQD for these states, we exploit the lower bound of GQD to this aim.

The remainder of the paper is organized as follows: In Sec. 2, after a brief review of quantum discord and the lower bound of GQD, we investigate the dynamics of a V-type three-level atom as an open quantum system. In Sec. 3, the quantum discord protection protocol for a two-qutrit system, by using the method of weak measurement and its reversal is presented and it is examined for the Werner and Horodecki states. Finally, we summarize our research in Sec. 4.
\section{Basic theory}
\subsection{Quantum discord}
At the first time, quantum discord which is one of the interesting measures of quantum correlations beyond entanglement was defined by the notion of the mutual information \cite{discord}. In classical information theory, the mutual information measures correlation between two random variables of a classical system, $A$ and $B$
\begin{equation}
I(A:B)=H(A)+H(B)-H(A,B),
\end{equation}
where $H(A)=-\sum_{a}p_{a} \log_{2}{p_{a}}$ is known as the Shannon entropy of the random variable $A$. In addition, $H(A,B)$ shows the joint Shannon entropy of two variables $A$ and $B$. Another equivalent definition of the classical mutual information is given by
\begin{equation}\label{2}
J(A:B)=H(B)-H(B|A).
\end{equation}
Notic that $H(B|A)$ indicates the information gained about the subsystem B by measuring the subsystem A.
The mutual information in quantum theory is defined as
\begin{equation}
I(A:B)=S(A)+S(B)-S(A,B),
\end{equation}
where $S(A)=-Tr[\rho_{A} log_{2}\rho_{A}]$ stands for the von Neumann entropy of the subsystem $A$ with the reduced density operator $\rho_{A}=tr_{B}\rho$. Two definitions of the classical mutual information are the same but for quantum systems differ from each other and the difference introduces quantum discord. An extension  of  the classical conditional entropy is $S(B|\{\Pi_{A}\})$ which shows the von Neumann entropy of $B$ conditioned on the outcome of the measurement on $A$. The quantum version of Eq. (\ref{2}) is as
\begin{equation}
J(A:B)=S(B)-S(B|\{\Pi_{A}\}).
\end{equation}
Ollivier and Zurek called the difference between $I(A:B)$ and $J(A:B)$ quantum discord \cite{Zurek} as
\begin{equation}\label{3}
\Delta^{B|A}=I(A:B)-J(A:B).
\end{equation}
According to the above definition, quantum discord is a function of the applied measurement on the subsystem $A$. To make quantum discord independent of such measurements, Henderson and Vederal \cite{vederal} performed a minimization of Eq. (\ref{3}) over all von Neumann measurements and obtained the following formula for quantum discord
\begin{equation}\label{vederaldiscord}
\delta^{B|A}=\mathrm{min}_{\{\Pi_{A}\}}[I(A:B)-J(A:B)].
\end{equation}
However, in general, Eq. (\ref{vederaldiscord}) for evaluating quantum discord is very difficult from calculational point view. To overcome this difficulty, a measure for quantum discord named as geometric quantum discord (GQD) was introduced by Dacki\'{c} $et$ $al$. \cite{dackic}, as follows
\begin{equation}\label{GQD}
D(\rho)=\mathrm{min}_{\chi\in\Omega_{0}}\|\rho-\chi\|^{2}.
\end{equation}
Here, $\Omega_{0}$ shows the set of zero-discord states and $\|A-B\|^{2}=Tr(A-B)^{2}$  expresses the square of Hilbert-Schmidt norm of Hermitian operators. Using the Bloch representation
 \begin{equation}\label{GQD}
\rho=\frac{1}{4}\big(I_{2}\otimes I_{2}+\sum_{i=1}^{3}x_{i}\sigma_i\otimes I_{2}+ \sum_{j=1}^{3}y_{j}I_{2}\otimes \sigma_j+\sum_{i,j=1}^{3}t_{ij}\sigma_i\otimes\sigma_j\big),
\end{equation}
GQD can be evaluated for every general two-qubit state $\rho$ as
\begin{equation}\label{discord of two qubit}
D(\rho)=\frac{1}{4}(\|x\|^{2}+\|T\|^{2}-\mu_{max} ),
\end{equation}
where $x=(x_{1}, x_{2}, x_{3})^{\dagger}$ is a column vector, $\|x\|^2=x^{\dagger}x, T=(t_{ij})$ is the correlation matrix and $\mu_{max}$ shows the largest eigenvalue of the matrix $x x^{\dagger}+T T^{\dagger}$. \\We can extend this result for the general bipartite system belonging to $H^{a}\otimes H^{b}$ with $\mathrm{dim}(H^{a})=m$ and  $\mathrm{dim}(H^{b})=n$. The vector space of all linear operators on $H^{a}$ with the Hilbert-Schmidt inner product
\begin{equation}
\langle X|Y\rangle=Tr(X^{\dagger}Y),
\end{equation}
constitutes a Hilbert space $L(H^{a} )$ (as well as $L(H^{b} )$ and $L(H^{a}\otimes H^{b})$). Two sets of Hermitian operators $\{X_{i}:i=1,2,\cdot\cdot\cdot,m^{2}\}$ and $\{Y_{j}: j=1,2,\cdot\cdot\cdot,n^{2}\}$ which satisfy the orthonormality conditions
\begin{equation}
Tr X_{i} X_{i^\prime}=\delta_{ii^{\prime}},\qquad   Tr Y_{j} Y_{j^\prime}=\delta_{jj^{\prime}},
\end{equation}
can be choosen to construct orthonormal bases for $L(H^{a})$  and $L(H^{b})$, respectively.
Then, $\{X_{i}\otimes Y_{j}\}$ is considered as an orthonormal basis for $L(H^{a}\otimes H^{b})$. The expansion of an arbitrary bipartite density matrix $\rho$ on $H^{a}\otimes H^{b}$ can be represented by
\begin{equation}
\rho=\sum_{ij}c_{ij} X_{i}\otimes Y_{j},
\end{equation}
where $c_{ij}= Tr \rho X_{i}\otimes Y_{j} $.
\\Let us consider  $\lambda_{i}$ ($i= 1, 2, \cdot\cdot\cdot, m^{2}-1$) and $\lambda_j$, $(j=1, 2, \cdot\cdot\cdot, n^{2}-1)$ each of which satisfy $Tr(\lambda_{i} \lambda_{j})=2\delta_{ij}$ as the generators of $SU(m)$ and $SU(n)$, respectively. By using this set of basis we can rewrite the bipartite density matrix $\rho$ on $H^{a}\otimes H^{b}$ as
\begin{equation}\label{bloch}
\rho=\frac{1}{mn}(I_{m}\otimes I_{n}+\sum_{i}x_{i}\lambda_i\otimes I_{n}+ \sum_{j}y_{j}I_{m}\otimes \lambda_j+\sum_{i,j}t_{ij}\lambda_i\otimes\lambda_j),
\end{equation}
where the $x_{i}$s and $y_{j}$s are the elements of the vectors $\vec{x}\in R^{m^{2}-1}$ and $\vec{y}\in R^{n^{2}-1}$  corresponding to
the subsystems $A$ and $B$ respectively obtained as
\begin{equation}
x_{i}=\frac{m}{2} Tr(\rho\lambda_{i}\otimes I_{n} )=\frac{m}{2} Tr(\rho_{A}\lambda_{i}),
\end{equation}
\begin{equation}
y_{j}=\frac{n}{2} Tr(\rho I_{m}\otimes\lambda_{j} )=\frac{n}{2} Tr(\rho_{B}\lambda_{j}).
\end{equation}
Note that
\begin{equation}
T=\frac{mn}{4}[Tr(\rho\lambda_{i}\otimes\lambda_{j})],
\end{equation}
is the correlation matrix then we have
\begin{equation}
C=\left(
\begin{array}{cc}
 \frac{I}{\sqrt{mn}} &\frac{\sqrt{2}}{n\sqrt{m}}{\vec{y}}^{\dagger} \\
 \frac{\sqrt{2}}{m\sqrt{n}}{\vec{x}}&\frac{2}{mn}T
\end{array}
\right).
\end{equation}
This matrix can also be written in terms of the basis $\{X_{i}\}$ and $\{Y_{j}\}$ as $C = [c_{ij} ] = Tr(\rho X_{i}\otimes Y_{j})$.
Since evaluation of $D(\rho)$ in Eq. (2.9) is a difficult procedure for high dimensional bipartite systems ($m,n\geq3$), we can use the lower bound of GQD introduced by Lu $et$ $al$. \cite{lowerbound} as
\begin{equation}\label{lowerboundofdiscord}
 D(\rho)\geq Tr(CC^{\dagger})-\sum_{i=1}^{m}\mu_{i}=\sum_{i=m+1}^{m^{2}}\mu_{i},
\end{equation}
instead of $D(\rho)$ in which $\mu_{i}$s show the eigenvalues of $CC^{\dagger}$ sorted in non-increasing order. So in this way, the lower bound of GQD in Eq. (2.18) is used as GQD in Sec. 3.
\subsection{Dynamics of V-type three-level open systems}
This section is devoted to investigate the interaction between a V-type three-level atom and a zero-temperature bosonic Lorentzian reservoir. In this configuration of the three-level atom, each of the excited levels $|2\rangle$ and $|1\rangle$ can decay to the ground state $|0\rangle$ and emit a single photon but the transition between $|2\rangle$ and $|1\rangle$ is dipole-forbidden.
The total Hamiltonian for the three-level system is composed of three terms
\begin{equation}
H=H_{S}+H_{E}+H_{I},
\end{equation}
where the first and second terms represent the free hamiltonian
\begin{equation}
H_{0}=H_{S}+H_{E}=\sum_{l=1}^{2}\omega_{l}\sigma_{+}^{l}\sigma_{-}^{l}+\sum_{k}\omega_{k}b_{k}^{\dagger}b_{k},
\end{equation}
and the third term is responsible for the atom-environment interaction Hamiltonian
\begin{equation}\label{interactionhamiltonian}
 H_{I}=\sum_{l=1}^{2}\sum_{k}(g_{lk}\sigma_{+}^{l}b_{k}+g_{lk}^{\ast}\sigma_{-}^{l}b_{k}^{\dagger}).
\end{equation}
In Eq. (\ref{interactionhamiltonian}), $\sigma_{+}^{l}=|l\rangle\langle0|$ and $\sigma_{-}^{l}=|0\rangle\langle l|\,(l=1, 2)$  indicate  atomic transition operators and $b_{k}$, $b_{k}^{\dagger}$ correspond to
operators of the field mode $k$ with the frequency $\omega_{k}$ and the atom-field coupling constant $g_{k}$. It is convenient to investigate the atom-field interaction problem in the interaction picture. In the interaction picture, the interaction term in Eq. (\ref{interactionhamiltonian}) is given by
\begin{equation}
H_{int}=\sum_{l=1}^{2}\sum_{k}(g_{lk}\sigma_{+}^{l}b_{k}e^{i(\omega_{l}-\omega_{k})t}+g_{lk}^{\ast}\sigma_{-}^{l}b_{k}^{\dagger}e^{-i(\omega_{l}-\omega_{k})t}),
\end{equation}
and the time-evolution of the total system is determined by the schr\"{o}dinger equation
\begin{equation}
i\frac{d}{dt}|\psi(t)\rangle=H_{int}|\psi(t)\rangle.
\end{equation}
At any time $t$, $|\psi(t)\rangle$ is expressed as a linear combination of the
states $|l\rangle_{S}\otimes|0\rangle_{E}$ and $|0\rangle_{S}\otimes|1_{k}\rangle_{E}$. Here $|l\rangle_{S}\otimes|0\rangle_{E}$ indicates that the atom is in the excited level  and the environment has no photon. A similar description exists for the state $|0\rangle_{S}\otimes|1_{k}\rangle_{E}$; therefore the state vector is
\begin{equation}\label{state vectore}
|\psi(t)\rangle=\sum_{l=0}^{2}c_{l}(t)|l\rangle_{S}\otimes|0\rangle_{E}+\sum_{k}c_{k}(t)|0\rangle_{S}\otimes|1_{k}\rangle_{E}.
\end{equation}
Using the schr\"{o}dinger equation, the time evaluation of probability
amplitudes $c_{1}(t)$ and $c_{2}(t)$  is governed
by the following coupled differential equations
\begin{equation}\label{d1}
\dot{c}_{l}(t)=-i\sum_{k}g_{lk}c_{k}(t)e^{i(\omega_{l}-\omega_{k})t},\quad (l=1, 2)
\end{equation}
\begin{equation}\label{d2}
 \dot{c}_{k}(t)=-i\sum_{l=1}^{2}g_{lk}^{\ast}c_{l}(t)e^{-i(\omega_{l}-\omega_{k})t}.
\end{equation}
Then substituting $c_{k}(t)$ from the soulution of Eq. (\ref{d2}) into Eq. (\ref{d1}) with the initial condition $c_{k}(0)=0$, i.e. we have no photon in the initial state of the environment, the time-varying probability amplitudes can be obtained as
\begin{equation}\label{amplitud1}
\dot{c_{l}}(t)=-\sum_{m=1}^{2}\int_{0}^{t}f_{lm}(t-t^{\prime})c_{m}(t^{\prime})dt^{\prime},\quad (l=1,2)
\end{equation}
where
\begin{equation}\label{amplitud2}
f_{lm}(t-t^{\prime})=\int_{0}^{t}d\omega J_{lm}(\omega)e^{i(\omega_{l}-\omega)t-i(\omega_{m}-\omega)t^{\prime}}.
\end{equation}
It should be noted that $c_{0}(t)$ remains unchanged because $H_{int}|0\rangle_{S}\otimes|0\rangle_{E}=0$. In Eq. (\ref{amplitud2}) $J_{lm}(\omega)$ denote the spectral density of the reservoir given by the following Lorentzian function
\begin{equation}
 J_{lm}(\omega)=\frac{1}{2\pi}\frac{\gamma_{lm}\lambda^{2}}{(\omega_{0}-\Delta-\omega)^{2}+\lambda^{2}},
\end{equation}
where $\Delta$ is defined as the detuning between the atomic transition frequency and the central frequency of the reservoir and $\lambda$ shows the spectral width of the coupling. Here $\gamma_{lm}=\sqrt{\gamma_{l}\gamma_{m}}\theta \quad(l\neq m$  and  $|\theta|\leq1)$ causes the spontaneously generated interference (SGI) between  the two decay channels $|2\rangle\longrightarrow|0\rangle$ and $|1\rangle\longrightarrow|0\rangle$ with $\theta$ denoting the relative angle between two dipole moment elements of these decay channels and $\gamma_{ll}=\gamma_{l}$ corresponds to the relaxation rate of the $l^{th}$ excited state. For $\theta=0$, dipole moment elements of the two decay channels are perpendicular  to each other and we do not have any SGI between them while $\theta=\pm1$ corresponds to the case of two parallel (or antiparallel) decay channels and shows the strongest SGI. Let $\tilde{c}_{l}(p)=L[c_{l}(t)]=\int_{0}^{\infty}c_{l}(t)e^{-pt}dt\: (l=1, 2)$, be the Laplace transform of $c_{l}(t)$. Taking the Laplace transform of Eq. (\ref{amplitud1}), we have
\begin{equation}\label{laplac}
  \left(
  \begin{array}{c}
  p\tilde{c}_{2}(p)-c_{2} (0)\\
  p\tilde{c}_{1}(p)-c_{1} (0)
\end{array}
\right)
  =-\frac{\lambda}{2(p+\lambda)}\left(
\begin{array}{cc}
  \gamma_{2} & -\sqrt{\gamma_{1}\gamma_{2}}\theta \\
  \sqrt{\gamma_{1}\gamma_{2}}\theta & \gamma_{1}
\end{array}
\right)
\left(
  \begin{array}{ccc}
  \tilde{c}_{2}(p)\\
  \tilde{c}_{1}(p)
\end{array}
\right).
\end{equation}
We need to apply the following unitary transformation on the Eq. \eqref{laplac}
\begin{equation}\label{unitary}
  \upsilon=\left(
  \begin{array}{cc}
             \sqrt{\frac{h+\gamma_{1}-\gamma_{2}}{2h}} &-\sqrt{\frac{h-\gamma_{1}+\gamma_{2}}{2h}} \\
             \sqrt{\frac{h-\gamma_{1}+\gamma_{2}}{2h}} & \sqrt{\frac{h+\gamma_{1}-\gamma_{2}}{2h}}
           \end{array}
           \right),
\end{equation}
and perform the inverse Laplace transform. Consequently, it is obtained
 \begin{equation}\label{cpm}
  c_{\pm}(t)=G_{\pm}(t)c_{\pm}(0),
 \end{equation}
in which $c_{\pm}(0)=\frac{1}{\sqrt{2h}}\big(c_{2}(0)\sqrt{h\pm\gamma_{1}\mp\gamma_{2}}\mp c_{1}(0)\sqrt{h\mp\gamma_{1}\pm\gamma_{2}}\big)$ are the initial probability amplitudes with  $h=\sqrt{(\gamma_{1}-\gamma_{2})^{2}+4\gamma_{1}\gamma_{2}\theta^{2}}$ and
 \begin{equation}\label{G+-}
 G_{\pm}(t)=e^{\frac{-\lambda t}{2}}\big(\cosh(\frac{d_{\pm}t}{2})+\frac{\lambda}{d_{\pm}}sinh(\frac{d_{\pm}t}{2})\big).
 \end{equation}
In Eq. (\ref{G+-}), $d_{\pm}=\sqrt{\lambda^{2}-2\lambda\gamma_{\pm}}$ and $\gamma_{\pm}=\frac{\gamma_{1}+\gamma_{2}\pm h}{2}$.
Notice that the unitary transformation mentioned above corresponds to the space spanned by $\{|2\rangle, |1\rangle\}$. It can be extended on the whole space of the system spanned by $\{|2\rangle, |1\rangle, |0\rangle\}$ as
\begin{equation}\label{unitarytransformation1}
 U=\left(
   \begin{array}{ccc}
     \sqrt{\frac{h+\gamma_{1}-\gamma_{2}}{2h}} & -\sqrt{\frac{h-\gamma_{1}+\gamma_{2}}{2h}}  & 0 \\
     \sqrt{\frac{h-\gamma_{1}+\gamma_{2}}{2h}} &  \sqrt{\frac{h+\gamma_{1}-\gamma_{2}}{2h}} & 0 \\
    0 & 0 & 1
   \end{array}
   \right).
 \end{equation}
With these assumptions, the density matrix of a three-level V-type atom at time $t$ takes the form
\begin{equation}\label{densitymatrix}
  \varrho_{s}(t)=\left(
  \begin{array}{ccc}
    |G_{+}(t)|^{2}|c_{+}(0)|^{2} & G_{+}(t)c_{+}(0)G_{-}^{\ast}(t)c_{-}^{\ast}(0) & G_{+}(t)c_{+}(0)c_{0}^{\ast} \\
    G_{+}^{\ast}(t)c_{+}^{\ast}(0)G_{-}(t)c_{-}(0) & |G_{-}(t)|^{2}|c_{-}(0)|^{2} &G_{-}(t)c_{-}(0)c_{0}^{\ast} \\
    G_{+}^{\ast}(t)c_{+}^{\ast}(0)c_{0} & G_{-}^{\ast}(t)c_{-}^{\ast}(0)c_{0} &1- |G_{+}(t)|^{2}|c_{+}(0)|^{2} -|G_{-}(t)|^{2}|c_{-}(0)|^{2}
  \end{array}
  \right),
 \end{equation}
which corresponds to  the time development of the state $\varrho_{s}(0)=U|\psi(0)\rangle\langle \psi(0)|U^{\dagger}$.
Here $|\psi(0)\rangle$ shows the state vector of the total system at time $t=0$.  \\
Let us introduce Kraus operators as
\begin{equation}
 \kappa_{1}=\left(
 \begin{array}{ccc}
   G_{+}(t) & 0 & 0 \\
   0 & G_{-}(t) & 0 \\
   0 & 0 & 1
 \end{array}
 \right),
 \end{equation}

 \begin{equation}
 \kappa_{2}=\left(
 \begin{array}{ccc}
   0 & 0 & 0 \\
   0 & 0 & 0 \\
   \sqrt{1-|G_{+}(t)|^{2}} & 0 & 0
 \end{array}
 \right),
 \end{equation}

 \begin{equation}
 \kappa_{3}=\left(
 \begin{array}{ccc}
   0 & 0 & 0 \\
   0 & 0 & 0 \\
   0&\sqrt{1-|G_{-}(t)|^{2}} & 0
 \end{array}
 \right).
 \end{equation}
It is convenient and useful to rewrite $\varrho_{s}(t)$  in terms of Kraus operators
\begin{equation}
\varrho_{s}(t)=\sum_{i=1}^{3}\kappa_{i}\varrho_{s}(0)\kappa_{i}^{\dagger},
\end{equation}
which satisfy the condition $\sum_{i=1}^{3}\kappa_{i}^{\dagger}\kappa_{i}=I_{3}$.
Finally, the original density matrix is given by
\begin{equation}
\rho_{s}(t)=U^{\dagger}\varrho_{s}(t)U.
\end{equation}
The above results can be applied to the case of a system involves two three-level V-type atoms
each of which couples independently to a Lorentzian reservoir. For such a two-qutrit system, the original density matrix at time $t$ is defined as
\begin{equation}
\rho_{s}(t)=(U\otimes U)^{\dagger}\varrho_{s}(t)(U\otimes U),
\end{equation}
where
\begin{equation}\label{twoqutrit}
\varrho_{s}(t)=\sum_{k,l=1}^{3}\kappa_{k,l}\varrho_{s}(0)\kappa_{k,l}^{^{\dagger}},
\end{equation}
with $\kappa_{k,l}=\kappa_{k}\otimes \kappa_{l}$ and $\sum_{k,l=1}^{3}\kappa_{k,l}^{\dagger}\kappa_{k,l}=I_{3}\otimes I_{3}$.

\section{Quantum discord protection}
In this section, based on WMQMR, we present a decoherence suppression scheme to protect GQD from the decoherence due to the AD channel.
We consider an arbitrary density matrix
$\rho_{s}(0)$ of the qutrit as the initial state
\begin{equation}
\rho_{s}(0)=\left(
  \begin{array}{ccc}
    a & b & c \\
    b^{\ast} & d & e \\
    c^{\ast} & e^{\ast} & f
  \end{array}
  \right),
\end{equation}
where $a+d+f=1$. Notice that $\rho_{s}(0)$ can represent the density matrix of a pure state with the state vector $|\psi(0)\rangle$ or a mixed state.\\
In the first step, we perform the unitary transformation of Eq. (\ref{unitarytransformation1}) with $\gamma_{1}=\gamma_{2}=\gamma$ on the initial density matrix of the qutrit
\begin{equation}
\varrho_{s}(0)=U\rho_{s}(0)U^{\dagger}.
\end{equation}
Next, a non-unitary operation as a prior weak measurement, $M_{w}$, is applied on the qutrit before it suffers decoherence arisen from the AD noise
\begin{equation}
\varrho_{s}^{w}(0)=M_{w}\varrho_{s}(0)M_{w}^{\dagger},
\end{equation}
with
\begin{equation}
  M_{w}=\left(
  \begin{array}{ccc}
    \sqrt{1-p} & 0 & 0 \\
   0 &\sqrt{1-q} & 0 \\
    0 & 0 & 1
  \end{array}
  \right),
\end{equation}
where $0\leq p,\: q < 1$ are called the weak measurement strengths which correspond to  transitions  $|2\rangle\longrightarrow|0\rangle$ and $|1\rangle\longrightarrow|0\rangle$, respectively.
In the next step, the state is altered by  the AD channel and the consequent state is given as
\begin{equation}
 \varrho_{s}^{w}(t)=\sum_{i=1}^{3}\kappa_{i}\varrho_{s}^{w}(0)\kappa_{i}^{\dagger}.
\end{equation}
According to the previous section, $\kappa_{i}\, (i=1,2,3)$  are  Kraus operators.
To recover the initial state, the system is subject to a post weak measurement reversal operation, $M_{r}$, which is also denoted by the following non-unitary operator
\begin{equation}
  M_{r}=\left(
\begin{array}{ccc}
    \sqrt{1-q_{r}} & 0 & 0 \\
   0 &\sqrt{1-p_{r}} & 0 \\
    0 & 0 & \sqrt{(1-q_{r})(1-p_{r})}
  \end{array}
  \right),
\end{equation}
where $ 0\leq p_{r}, \:q_{r} <1 $ indicate  strengths of the reversing measurement.
The density matrix of the qutrit with respect to this measurement is defined by
\begin{equation}\label{afterrevers}
 \varrho_{s}^{r}(t)=M_{r}\varrho_{s}^{w}(t)M_{r}^{\dagger}.
\end{equation}
It is demonstrated that $ p_{r}=1-(1-p)|G_{+}|^2$ and $q_{r}= 1-(1-q)|G_{-}|^2$ provide the best restoration of the origional state. Moreover, it should be noted that $G_{+}(t)$ and $G_{-}(t)$ are in fact
two complex functions, as a result we have
\begin{equation}
G_{+}(t)= |G_{+}|e^{i\varphi_{+}(t)},
\end{equation}
\begin{equation}
G_{-}(t)= |G_{-}|e^{i\varphi_{-}(t)}.
\end{equation}
Substituting $p_{r}$, $q_{r}$, $G_{+}(t)$ and $G_{-}(t)$ into the $\varrho_{s}^{r}(t)$, we obtain
\begin{equation}\label{optimizationmatrix}
\varrho_{s}^{r}(t)=R\left(
\begin{array}{ccc}
    \frac{1}{2}(a-b^{\ast}-b+d) & \frac{1}{2}(a-b^{\ast}+b-d) e^{i(\varphi_{+}(t)-\varphi_{-}(t))} & \frac{1}{\sqrt{2}}(c-e) e^{i\varphi_{+}(t)} \\
   \frac{1}{2}(a+b^{\ast}-b-d) e^{-i(\varphi_{+}(t)-\varphi_{-}(t))}&\frac{1}{2}(a+b^{\ast}+b+d) & \frac{1}{\sqrt{2}}(c+e) e^{i\varphi_{-}(t)} \\
   \frac{1}{\sqrt{2}}(c^{\ast}-e^{\ast}) e^{-i\varphi_{+}(t)} & \frac{1}{\sqrt{2}}(c^{\ast}+e^{\ast}) e^{-i\varphi_{-}(t)} &f+f^{\prime}
  \end{array}
  \right),
\end{equation}
with $R=(1-p)(1-q)|G_{+}|^{2}|G_{-}|^{2}$ and $f^{\prime}=\frac{1}{2}(a-b^{\ast}-b+d)(1-p)(1-|G_{+}|^{2})+\frac{1}{2}(a+b^{\ast}+b+d)(1-q)(1-|G_{-}|^{2})$. In order to achieve the initial state of the qutrit, first we have to use the matrix
\begin{equation}
V=\left(
\begin{array}{ccc}
  e^{i\varphi_{+}(t)} & 0 & 0 \\
  0 & e^{i\varphi_{-}(t)} & 0 \\
  0 & 0 & 1
\end{array}
\right),
\end{equation}
to omit  phase factors from Eq. (\ref{optimizationmatrix}) in the following way
\begin{equation}
\varrho_{s}(t)=V^{\dagger}\varrho _{s}^{r}(t)V.
\end{equation}
Then we need to apply the unitary transformation on $\varrho_{s}(t)$ as
\begin{equation}\label{finalstate}
\rho_{s}(t)=U^{\dagger}\varrho_{s}(t)U.
\end{equation}
Finally, the normalized qutrit state is described by
\begin{align}
\rho(t)=\frac{R}{N}(\rho_{s}(0)
+f^{\prime}
|0\rangle\langle0|),
\end{align}
where
\begin{align}
N=R(1+f^{\prime}),
\end{align}
represents the normalization factor.\\
It is easy to extend this scheme to a two-qutrit system. The efficiency of our geometric discord protection protocol  will be examined for two different families of two-qutrit systems such as Werner and Horodecki states. Let us begin with the Werner state.
\subsection{Werner state}
In this section, the initial state of the system is assumed to be prepared in a two-qutrit Werner state
\begin{equation}
\rho_{\eta}(0)=(1-\eta)\frac{I_{9}}{9}+\eta|\psi_{0}\rangle\langle\psi_{0}|,
\end{equation}
where $|\Psi_{0}\rangle=\frac{1}{\sqrt{3}}\big(|22\rangle+|11\rangle+|00\rangle\big)$, $I_{9}$ shows the identity matrix and $\eta \in[0,1]$ is called the purity parameter. Notice that for $\eta\leq\frac{1}{4}$, the Werner state is PPT, whereas it will be NPPT for $\eta > \frac{1}{4}$ \cite{werner}.
The density matrix of a two-qutrit Werner state under the protection protocol  discussed above is given by
\begin{align}
 \rho_{\eta}(t)=&\frac{R^{2}}{N_{w}}\bigg\{\rho_{\eta}(0)+\Big(\frac{\eta}{3}s_{1}+\frac{(1-\eta)}{9}(s_{2}+2s_{3})\Big)|00\rangle\langle00|\nonumber \\
 +&\frac{(\eta+2)}{18}s_{3}\Big(\,|20\rangle\langle20|+|02\rangle\langle02|+|10\rangle\langle10|+|01\rangle\langle01|\,\Big)\nonumber \\
 +&\frac{\eta}{6}s_{4}\Big(\,|10\rangle\langle20|+|01\rangle\langle02|+|20\rangle\langle10|+|02\rangle\langle01|\,\Big)\bigg\},
\end{align}
where
\begin{equation}
N_{w}=R^{2}\big(1+\frac{\eta}{3}s_{1}+\frac{(1-\eta)}{9}s_{2}+\frac{2}{3}s_{3}\big),
\end{equation}
is the normalization factor of $\rho_{\eta}(t)$
and
\begin{align}
&s_{1}=(1-p)^{2}(1-|G_{+}|^{2})^{2}+(1-q)^{2}(1-|G_{-}|^{2})^{2},
\end{align}
\begin{align}
s_{2}=s_{1}+2(1-p)(1-q)(1-|G_{+}|^{2})(1-|G_{-}|^{2}),
\end{align}
\begin{align}
&s_{3}=(1-p)(1-|G_{+}|^{2})+(1-q)(1-|G_{-}|^{2}),
\end{align}
\begin{align}
&s_{4}=-(1-p)(1-|G_{+}|^{2})+(1-q)(1-|G_{-}|^{2}).
\end{align}
Then using of Eq. \eqref{lowerboundofdiscord}, we can numerically calculate the lower bound of GQD for $\rho_{\eta}(t)$.
In Figs. $1, 2$ and $3$,  the variations of GQD versus $\gamma t$ and $\eta$ are plotted for different values of $\lambda$ and $p$ when the two-qutrit system is initially in the Werner state. It should be noted that, without lose of generality, $p=q$.
One can observe that for the case of $p=0$, the GQD is decreased with time (for both cases $\lambda=0.1$ and $\lambda=1$) although the decrement rate for $\lambda=0.1$ is smaller than the corresponding case of $\lambda=1$, as is evident from Figs. 1. Under this condition, the GQD, ultimately, approaches to zero. As the weak measurement strength is non-zero ($p=0.5$), the GQD can be protected against the dissipation but not completely (see Figs. 2). When the weak measurement strength $p\rightarrow 1$ (for example consider $p=0.99$), the two-qutrit GQD for the Werner state becomes to be well-protected against decoherence arisen from AD channel, as observed from Figs. 3.
It is interesting to note that in the case of $p\rightarrow 1$, the efficiency of protection protocol is independent of the values of $\lambda$. This, in fact, leads to provide the protection of the GQD for the two-qutrit Werner state from decoherence in both Markovian and non-Markovian dynamics.
\subsection{Horodecki state}
In the next step, to show further the efficiency of the protocol, we take the well-known two-qutrit Horodecki state \cite{Horodecki} as an initial state as follows
\begin{equation}
 \rho_{\alpha}(0)=\frac{2}{7}|\psi_{0}\rangle\langle\psi_{0}|+\frac{\alpha}{7}\sigma_{+}+\frac{5-\alpha}{7}\sigma_{-},
 \end{equation}
where $0\leq\alpha\leq5$ and
\begin{equation}
\sigma_{+}=\frac{1}{3}\big(|01\rangle\langle01|+|12\rangle\langle12|+|20\rangle\langle 20|\big),
\end{equation}
\begin{equation}
\sigma_{-}=\frac{1}{3}\big(|10\rangle\langle10|+|21\rangle\langle21|+|02\rangle\langle 02|\big).
\end{equation}
It is known that the Horodecki state remains unchanged with the replacement of $\alpha$ by $(5-\alpha)$. $\rho_{\alpha}(0)$ is (i) separable for $2\leq\alpha\leq3$, (ii) bound entangled for $1\leq\alpha<2$ and $3<\alpha\leq4$, (iii) free entangled for $0\leq\alpha<1$ and $4<\alpha\leq5$.
By considering the scheme for the Horodecki state, the final state after the related process becomes as
\begin{align}
 \rho_{\alpha}(t)=&\frac{R^{2}}{N_{\alpha}}\bigg\{\rho_{\alpha}(0)+\frac{1}{21}(2s_{1}+\frac{5}{4}s_{2}+5s_{3})|00\rangle\langle00|\nonumber \\
 +&\frac{1}{21}s_{3}\,\Big(\,|20\rangle\langle20|+|02\rangle\langle02|+|10\rangle\langle10|+|01\rangle\langle01|\,\Big)\nonumber \\
 +&\frac{1}{21}s_{4}\,\Big(\,|10\rangle\langle20|+|01\rangle\langle02|+|20\rangle\langle10|+|02\rangle\langle01|\,\Big)\nonumber\\
 +&\frac{\alpha}{42}s_{3}\,\Big(\,|20\rangle\langle20|+|01\rangle\langle01|-|10\rangle\langle10|-|02\rangle\langle02|\,\Big)\nonumber \\
 +&\frac{5}{42}s_{3}\,\Big(\,|10\rangle\langle10|+|02\rangle\langle02|\,\Big)\bigg\},
\end{align}
with the normalization factor
\begin{equation}
N_{\alpha}=R^{2}\big(1+\frac{1}{21}(2s_{1}+\frac{5}{4}s_{2}+14s_{3})\big).
\end{equation}
As it is expected, numerical results show that the increment of weak measurement strength $p$ leads to the improvement of the protection process of the quantum discord against the decoherence of AD channel, as shown in Figs. 4, 5 and 6. As for the case of Werner state, when $p=0$ the GQD suffers sudden death (especially for the Markovian dynamics corresponding to the Fig. 4b). Figs. 5, show that a non-zero $p$ provides the protection of the GQD from sudden death. Consequently, for $p\rightarrow1$, the GQD for the Horodecki state is also well-protected against the AD dissipation (see Figs. 6). It is worthwhile to mention that, apart from that the dynamical evolution of the two-qurit system is Markovian or non-Markovian, the initial time GQD of the Horodecki state can also be protected from the influences of AD channel.
\section{Conclusions}
In this work, we first studied the lower bound of GQD to quantify the two-qutrit quantum discord. In the next step, we
investigated the dynamics of a two-qutrit system each of which interacts with a dissipative environment independently. Then we showed that WMQMR can be used for protecting GQD of two-qutrit systems against the dissipative AD channel. To confirm the efficiency of our approach, we examined it for the two-qutrit Werner and Horodecki states.
It was found that for the strength of weak measurement with $p\rightarrow1$, the GQD could be well-preserved. In general, the efficiency of the proposed scheme is independent from weak or strong coupling of the system to the surrendering environment, i.e. Markovian or non-Markovian dynamics of the system.\\
\\


 \vspace{1cm}\setcounter{section}{0}
 \setcounter{equation}{0}
 \renewcommand{\theequation}{C-\roman{equation}}


\setlength\abovecaptionskip{0pt}
\begin{figure}[h]
\centering
\begin{subfigure}{0.6\linewidth}
  \centering
  \includegraphics[scale=0.5]{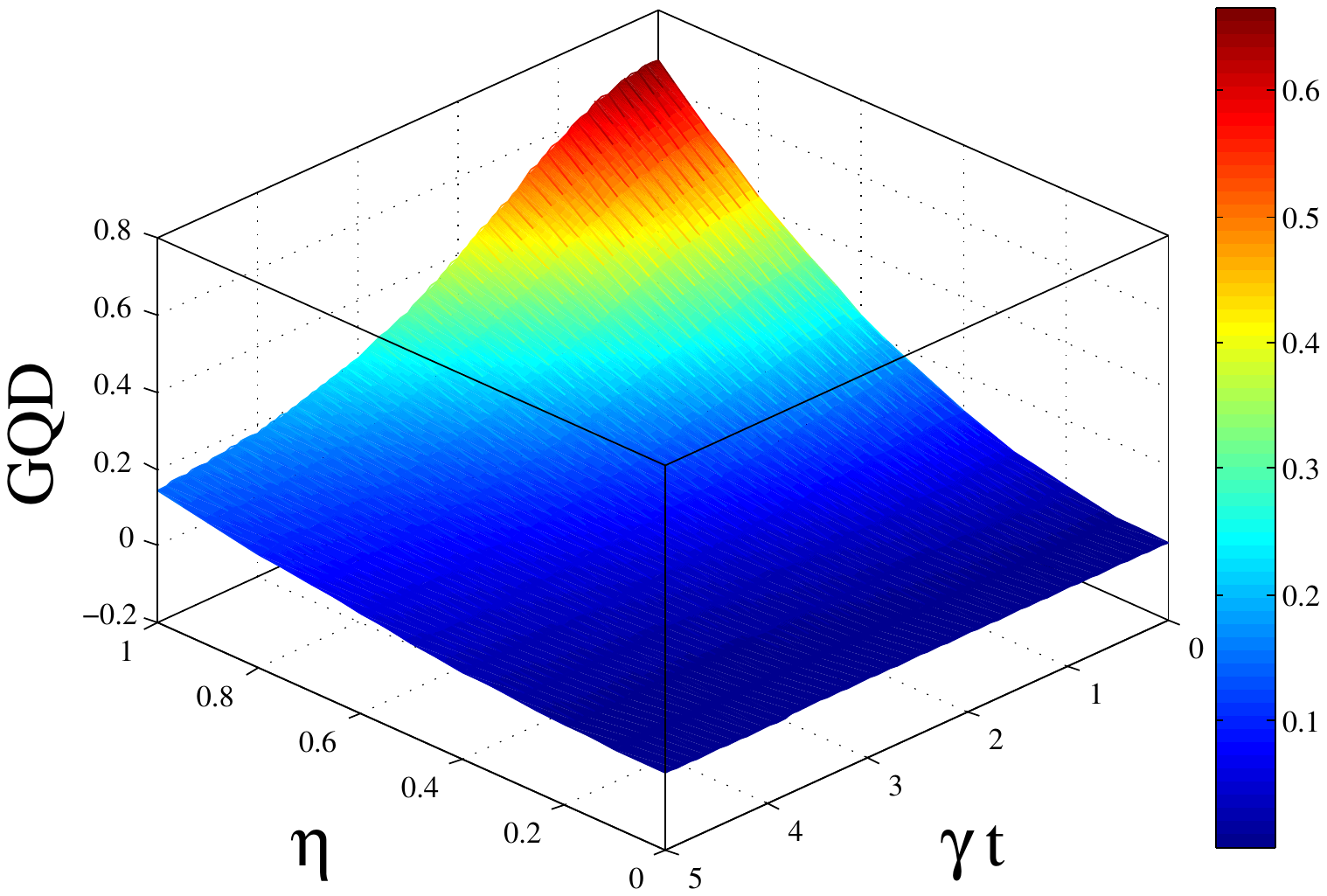}
  \vspace*{-6cm}
  \caption{}
\end{subfigure}%
\hspace{-3.30cm}
\begin{subfigure}{0.6\linewidth}
  \centering
  \includegraphics[scale=0.5]{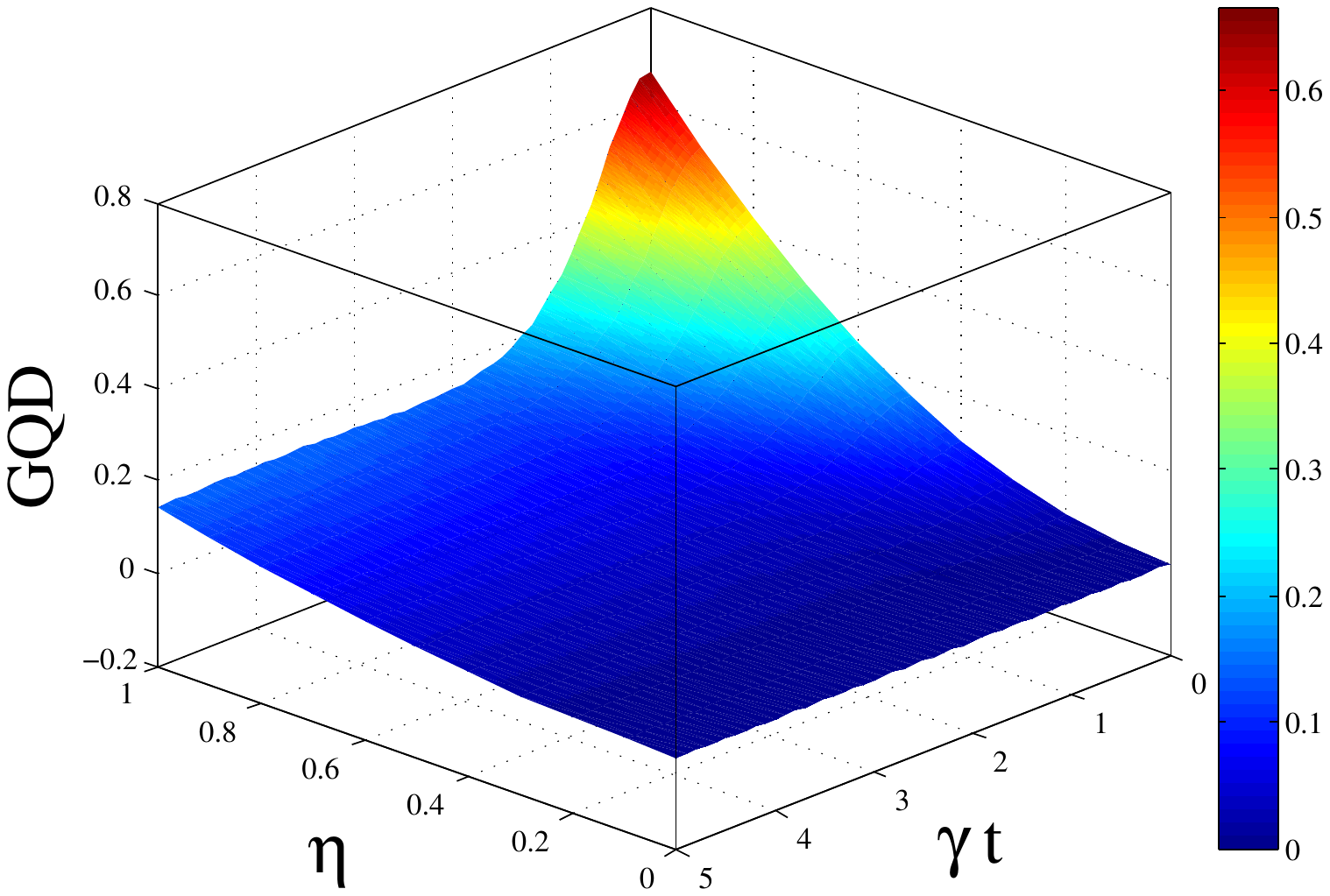}
  \vspace*{-6cm}
  \caption{}
\end{subfigure}
\caption{Geometric discord for the Werner state as functions of $\eta$ and $\gamma t$ without the protection protocol for (a) $\lambda=0.1$ and (b) $\lambda=1$, with $p=0$.}
\end{figure}
\setlength\abovecaptionskip{0pt}
\begin{figure}[h]
\centering
\begin{subfigure}{0.6\linewidth}
  \centering
  \includegraphics[scale=0.5]{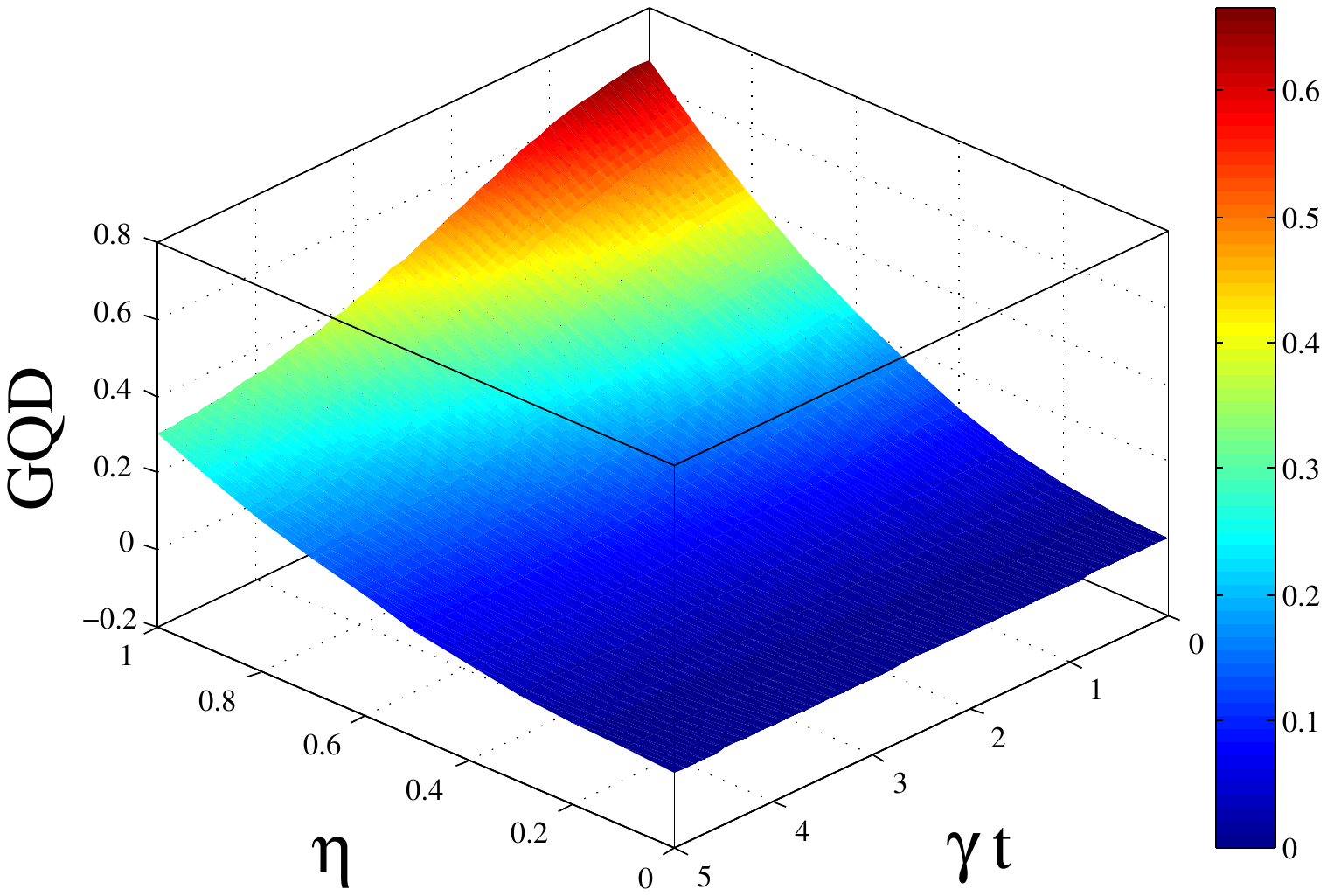}
  \vspace*{-6cm}
  \caption{}
\end{subfigure}%
\hspace{-3.30cm}
\begin{subfigure}{0.6\linewidth}
  \centering
  \includegraphics[scale=0.5]{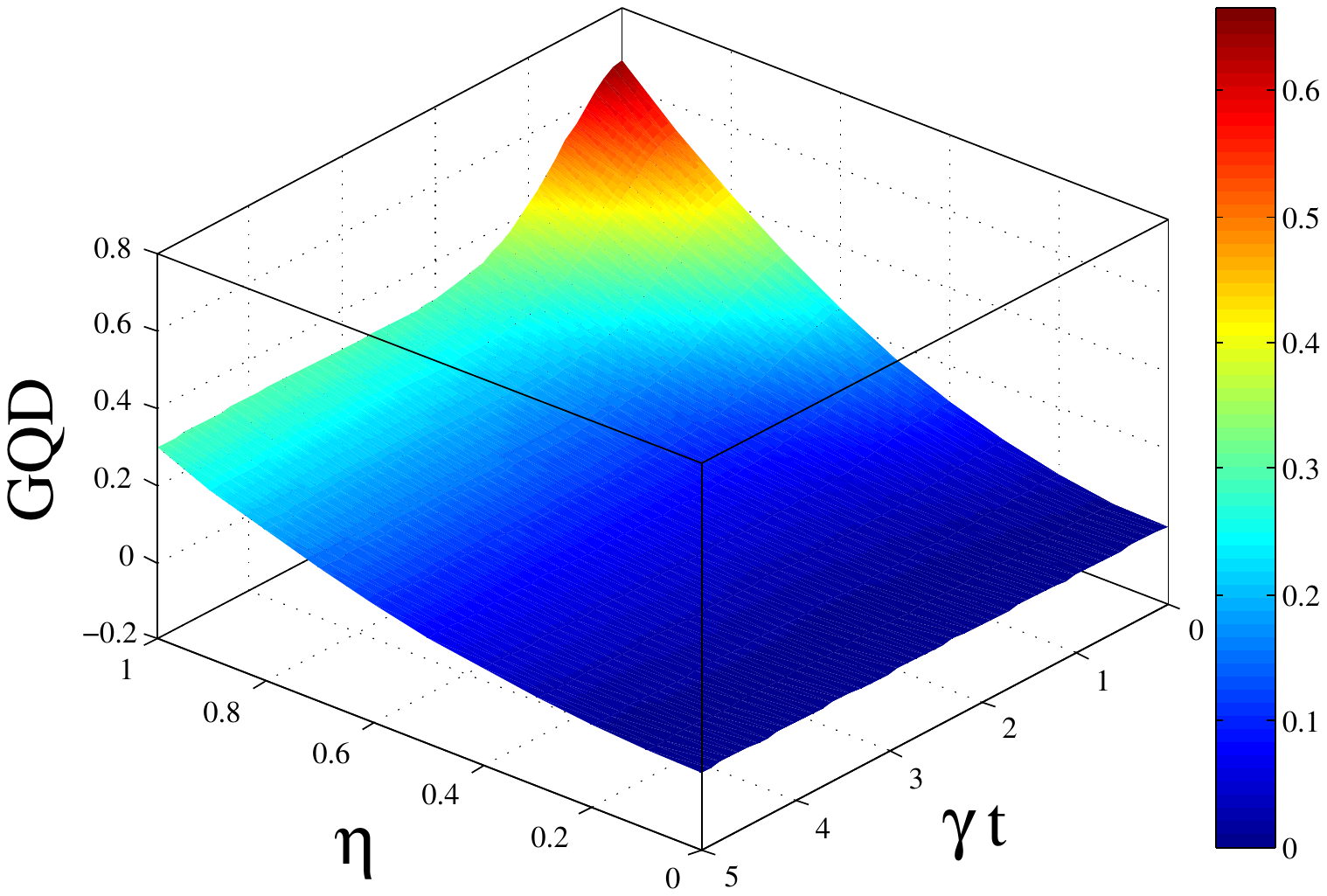}
  \vspace*{-6cm}
  \caption{}
\end{subfigure}
\caption{Geometric discord for the Werner state as functions of $\eta$ and $\gamma t$ under the protection protocol for (a) $\lambda=0.1$ and (b) $\lambda=1$, with $p=0.5$.}
\end{figure}
\setlength\abovecaptionskip{0pt}
\begin{figure}[h]
\centering
\begin{subfigure}{0.6\linewidth}
  \centering
  \includegraphics[scale=0.5]{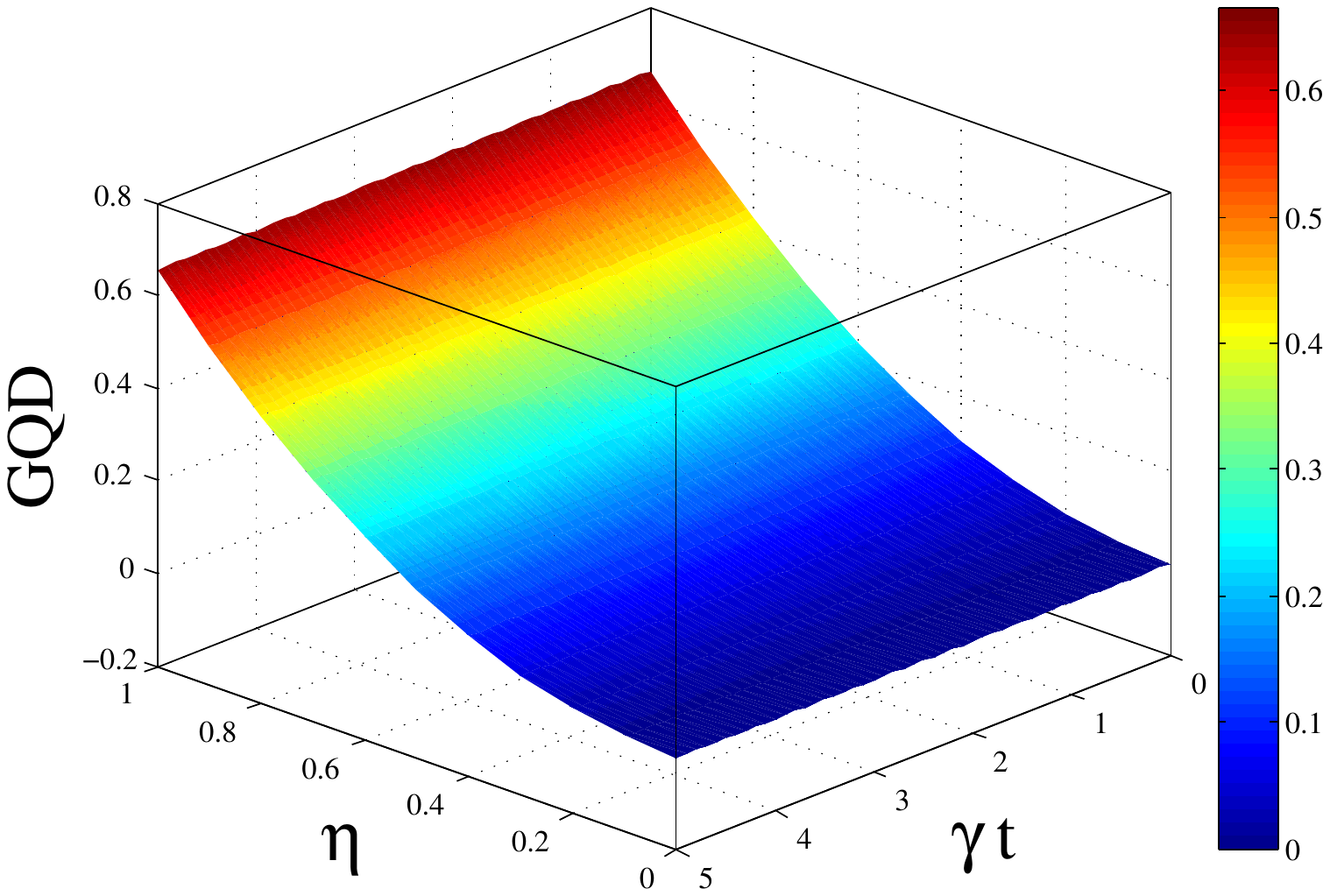}
  \vspace*{-6cm}
  \caption{}
\end{subfigure}%
\hspace{-3.30cm}
\begin{subfigure}{0.6\linewidth}
  \centering
  \includegraphics[scale=0.5]{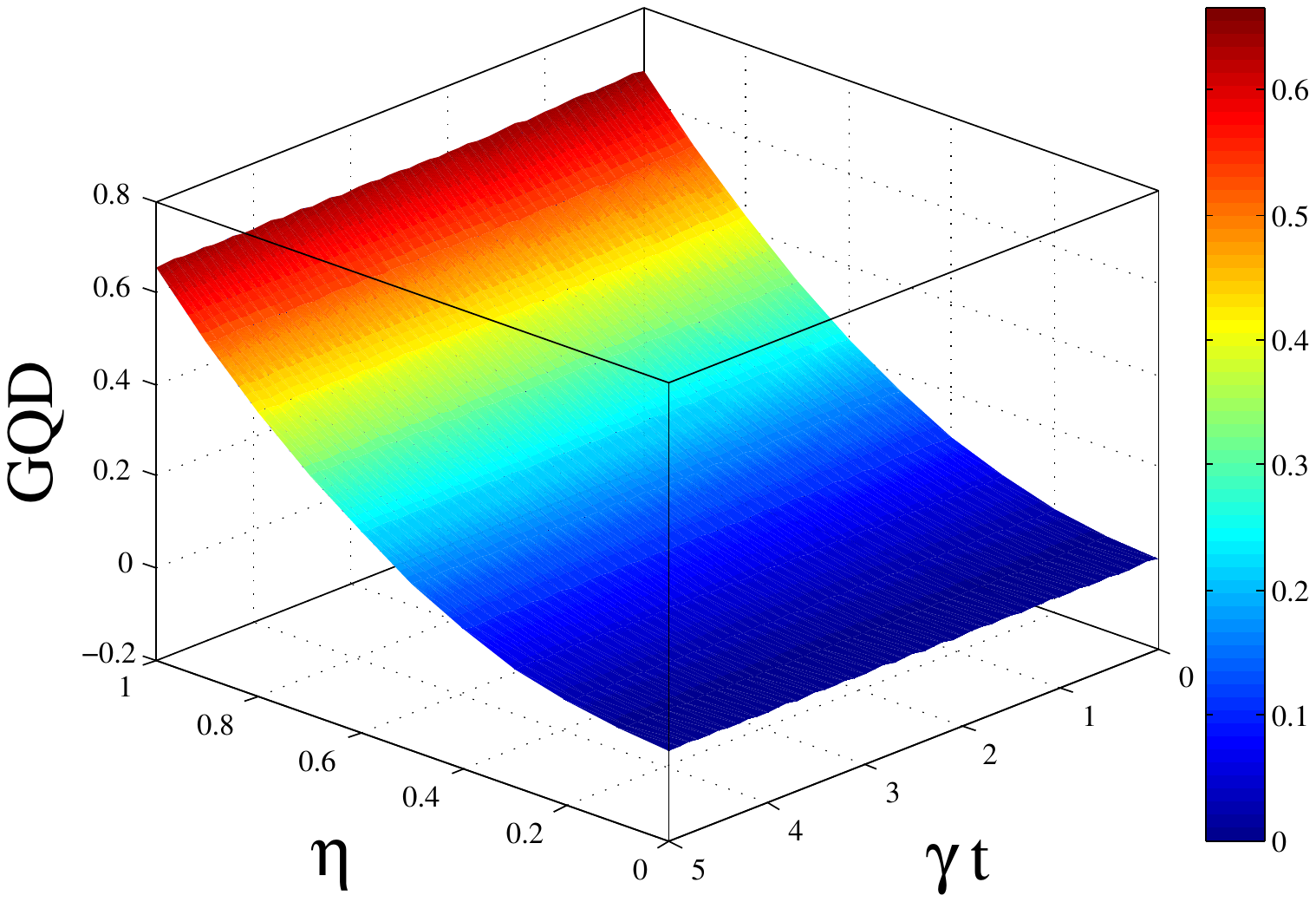}
  \vspace*{-6cm}
  \caption{}
\end{subfigure}
\caption{Geometric discord for the Werner state as functions of $\eta$ and $\gamma t$ under the protection protocol for (a) $\lambda=0.1$ and (b) $\lambda=1$, with $p=0.99$.}
\end{figure}
\setlength\abovecaptionskip{0pt}
\begin{figure}[h]
\centering
\begin{subfigure}{0.6\linewidth}
  \centering
  \includegraphics[scale=0.5]{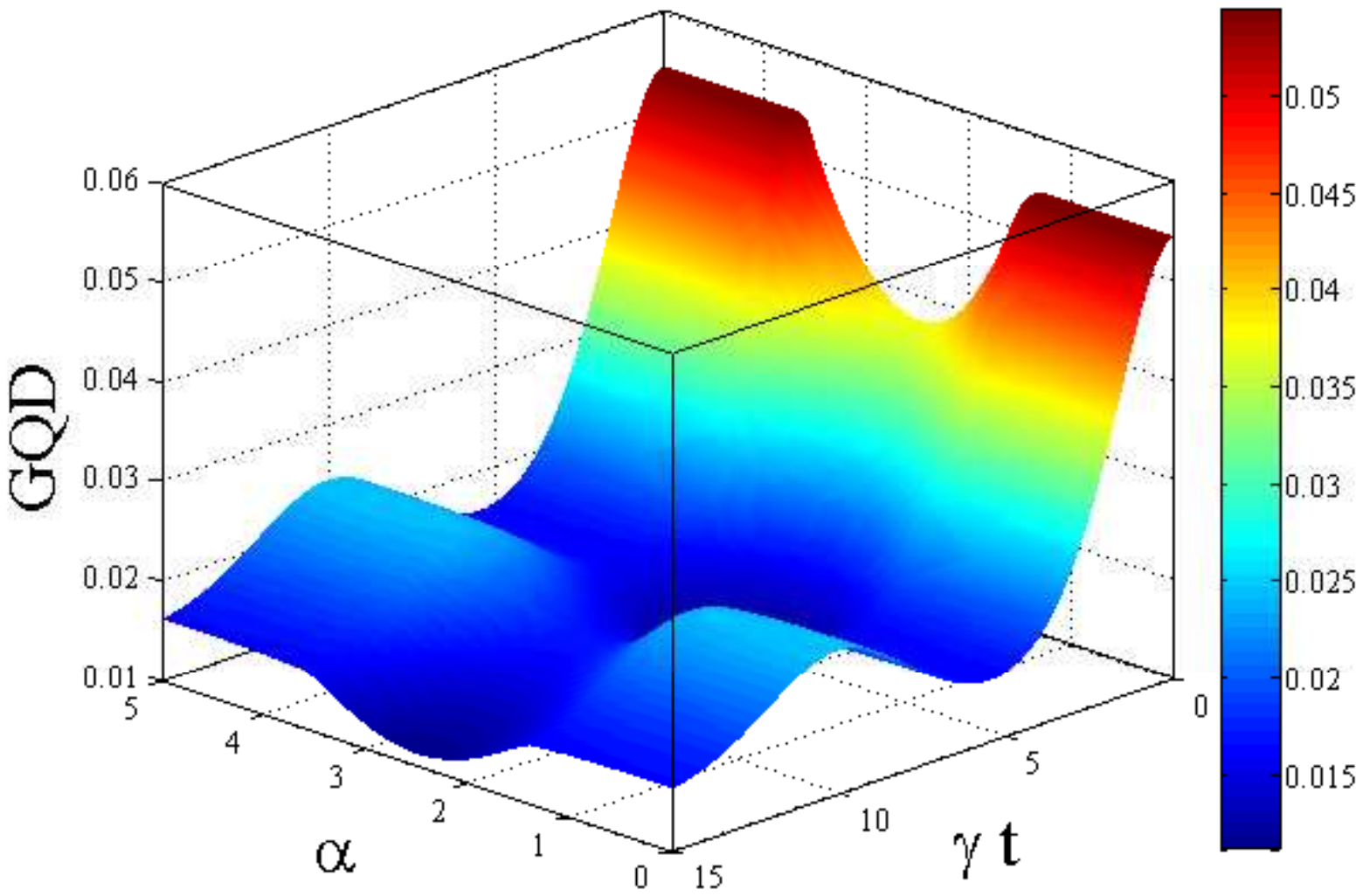}
  \vspace*{-6cm}
  \caption{}
\end{subfigure}%
\hspace{-3.30cm}
\begin{subfigure}{0.6\linewidth}
  \centering
  \includegraphics[scale=0.5]{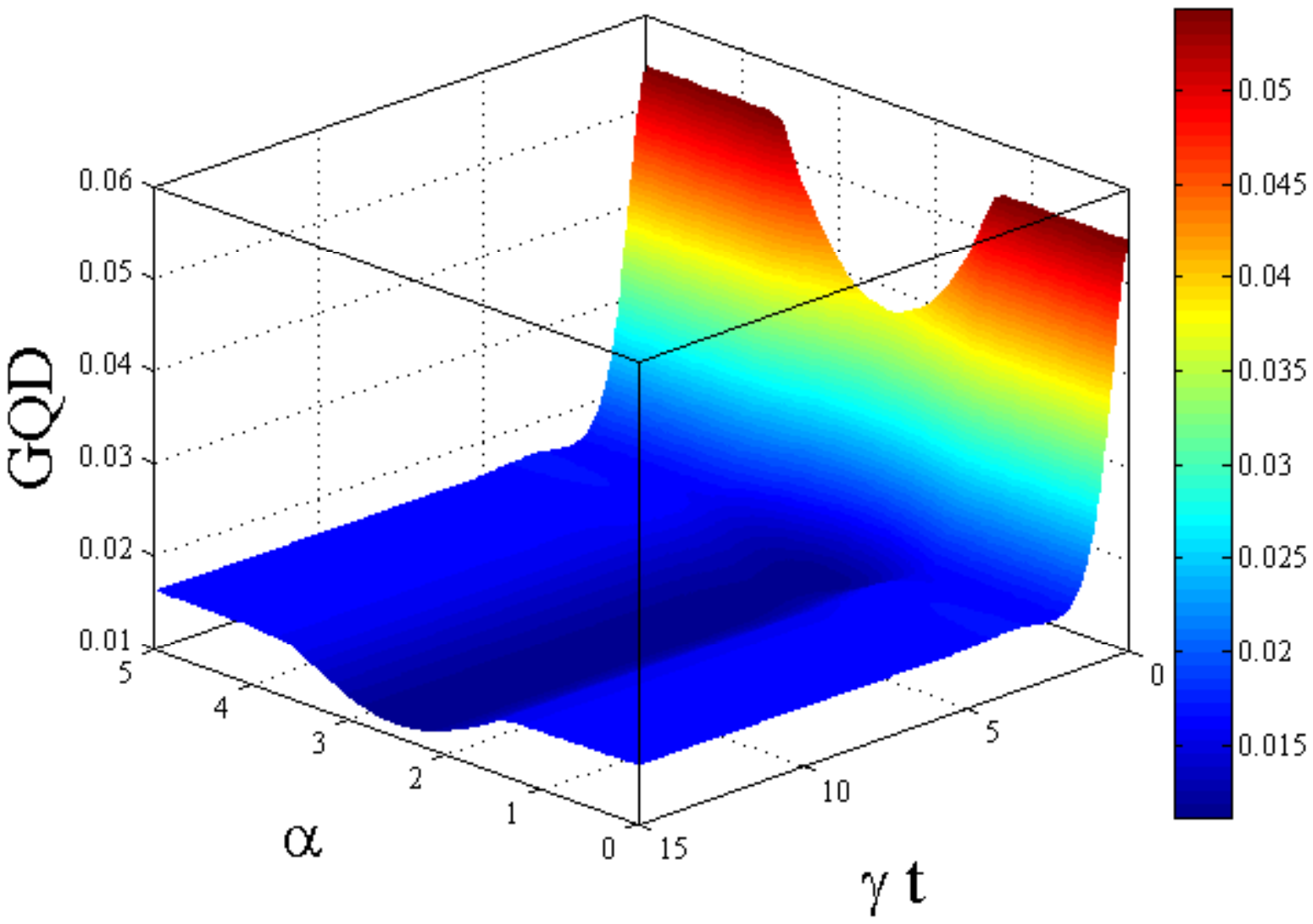}
  \vspace*{-6cm}
  \caption{}
\end{subfigure}
\caption{Geometric discord for the Horodecki state as functions of $\alpha$ and $\gamma t$ without the protection protocol for (a) $\lambda=0.1$ and (b) $\lambda=1$, with $p=0$.}
\end{figure}
\setlength\abovecaptionskip{0pt}
\begin{figure}[h]
\centering
\begin{subfigure}{0.6\linewidth}
  \centering
  \includegraphics[scale=0.5]{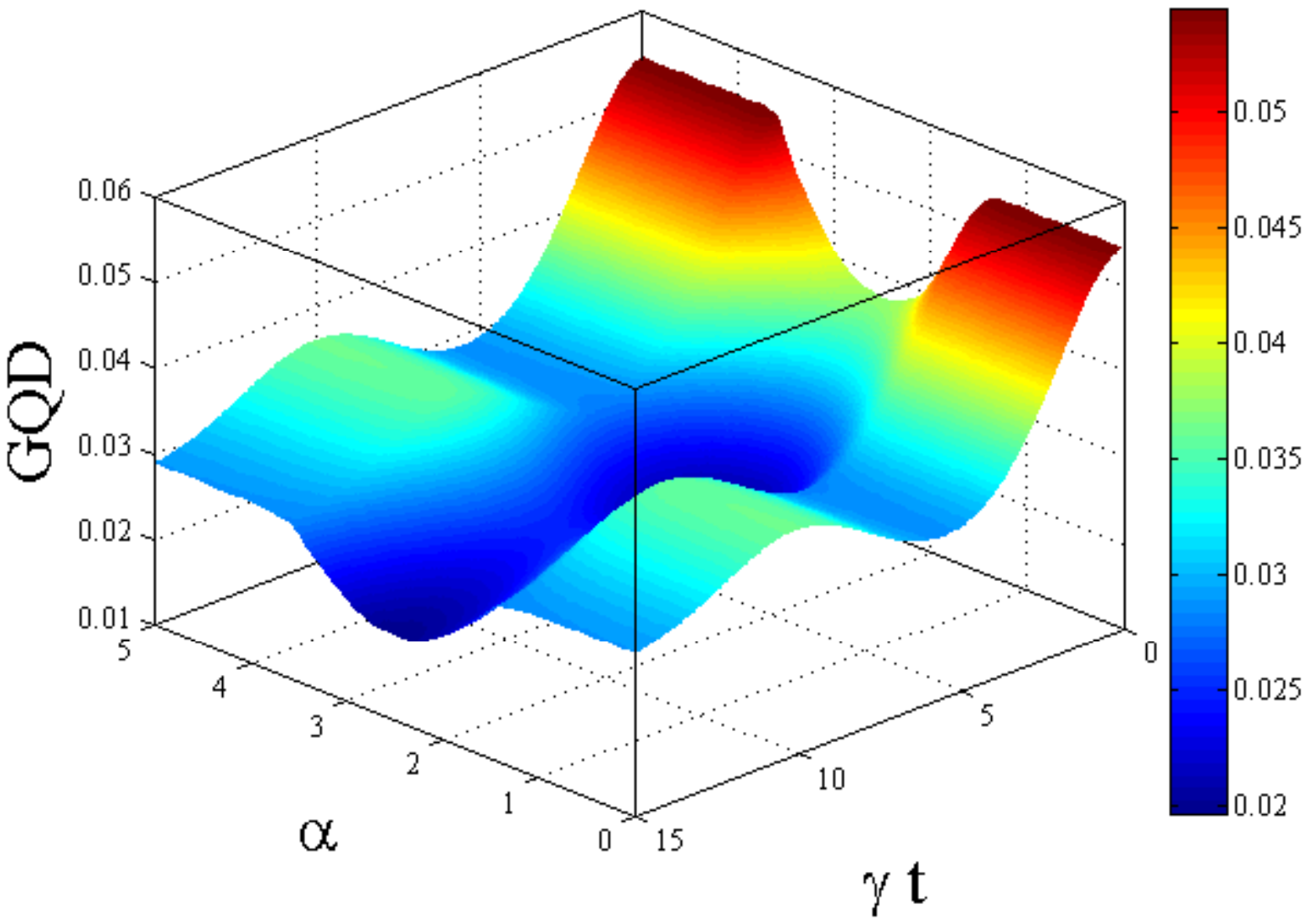}
  \vspace*{-6cm}
  \caption{}
\end{subfigure}%
\hspace{-3.30cm}
\begin{subfigure}{0.6\linewidth}
  \centering
  \includegraphics[scale=0.5]{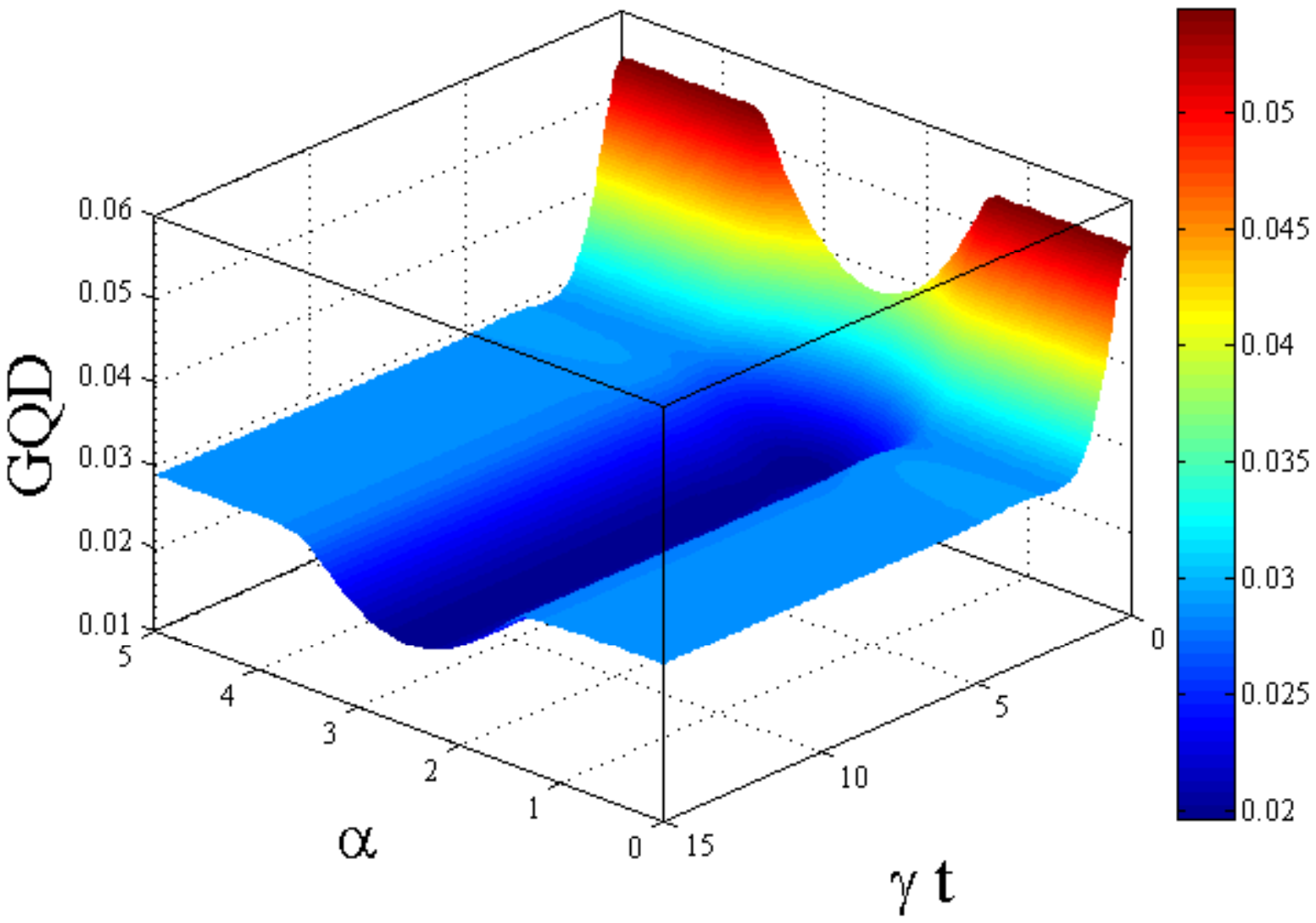}
  \vspace*{-6cm}
  \caption{}
\end{subfigure}
\caption{Geometric discord for the Horodecki state as functions of $\alpha$ and $\gamma t$ under the protection protocol for (a) $\lambda=0.1$ and (b) $\lambda=1$, with $p=0.5$.}
\end{figure}
\setlength\abovecaptionskip{0pt}
\begin{figure}[h]
\centering
\begin{subfigure}{0.6\linewidth}
  \centering
  \includegraphics[scale=0.5]{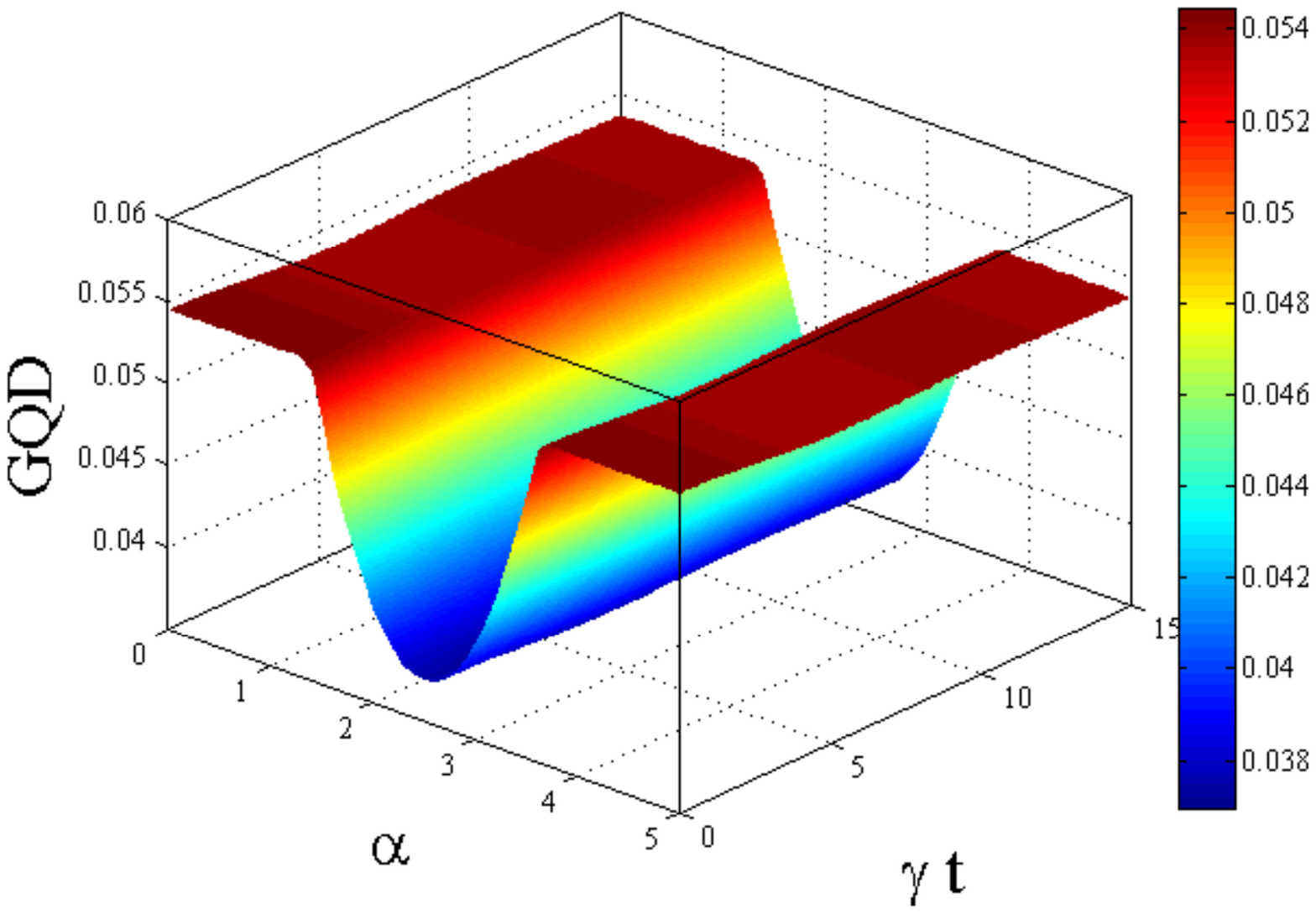}
  \vspace*{-6cm}
  \caption{}
\end{subfigure}%
\hspace{-3.30cm}
\begin{subfigure}{0.6\linewidth}
  \centering
  \includegraphics[scale=0.5]{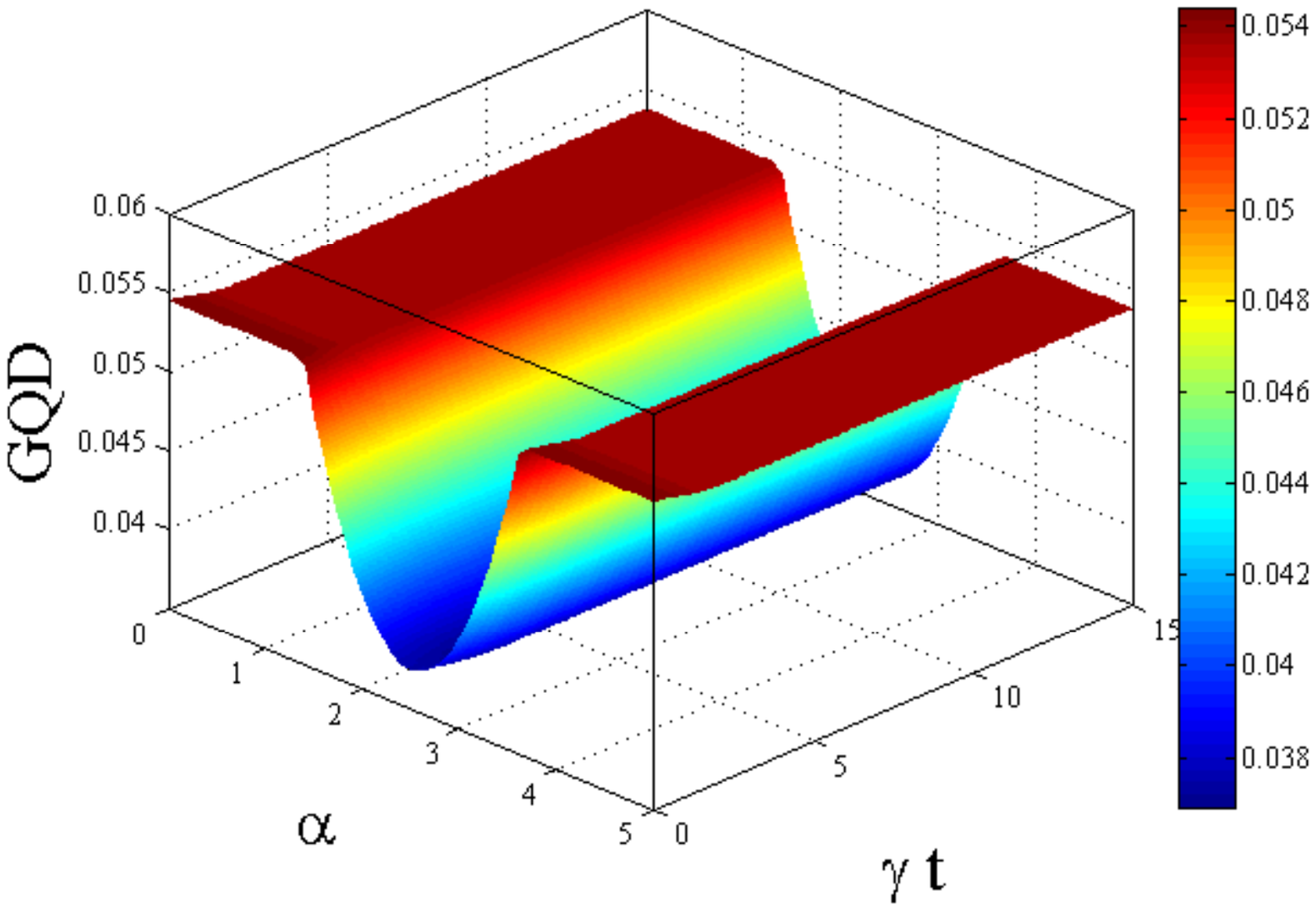}
  \vspace*{-6cm}
  \caption{}
\end{subfigure}
\caption{Geometric discord for the Horodecki state as functions of $\alpha$ and $\gamma t$ under the protection protocol for (a) $\lambda=0.1$ and (b) $\lambda=1$, with $p=0.99$.}
\end{figure}
\end{document}